\documentclass[aps,prl,preprint,nopacs,superscriptaddress]{revtex4}
\usepackage{graphicx}
\usepackage{verbatim}
\usepackage{mathrsfs}
\usepackage{amsmath,amsfonts,amssymb}
\usepackage{graphicx}

\def\3{2.8in}    
\def\2{2.5in}
\def\4{3.0in}

\def \beq {\begin{equation}}
\def \eeq {\end{equation}}
\begin{document}

\title{A new type of Weyl semimetal with quadratic double Weyl fermions in SrSi$_2$}

\author{Shin-Ming Huang\footnote{These authors contributed equally to this work.}}
\affiliation{Centre for Advanced 2D Materials and Graphene Research Centre National University of Singapore, 6 Science Drive 2, Singapore 117546 and
Department of Physics, National University of Singapore, 2 Science Drive 3, Singapore 117542}
\author{Su-Yang Xu$^*$}
\affiliation {Joseph Henry Laboratory, Department of Physics, Princeton University, Princeton, New Jersey 08544, USA}
\author{Ilya Belopolski}\affiliation {Joseph Henry Laboratory, Department of Physics, Princeton University, Princeton, New Jersey 08544, USA}
\author{Chi-Cheng Lee}
\affiliation{Centre for Advanced 2D Materials and Graphene Research Centre National University of Singapore, 6 Science Drive 2, Singapore 117546 and
Department of Physics, National University of Singapore, 2 Science Drive 3, Singapore 117542}
\author{Guoqing Chang}
\affiliation{Centre for Advanced 2D Materials and Graphene Research Centre National University of Singapore, 6 Science Drive 2, Singapore 117546 and
Department of Physics, National University of Singapore, 2 Science Drive 3, Singapore 117542}
\author{BaoKai Wang}
\affiliation{Centre for Advanced 2D Materials and Graphene Research Centre National University of Singapore, 6 Science Drive 2, Singapore 117546 and
Department of Physics, National University of Singapore, 2 Science Drive 3, Singapore 117542}
\affiliation{Department of Physics, Northeastern University, Boston, Massachusetts 02115, USA}
\author{Nasser Alidoust}\affiliation {Joseph Henry Laboratory, Department of Physics, Princeton University, Princeton, New Jersey 08544, USA}
\author{Madhab Neupane}\affiliation {Joseph Henry Laboratory, Department of Physics, Princeton University, Princeton, New Jersey 08544, USA}
\author{Hao Zheng}\affiliation {Joseph Henry Laboratory, Department of Physics, Princeton University, Princeton, New Jersey 08544, USA}
\author{Daniel Sanchez}\affiliation {Joseph Henry Laboratory, Department of Physics, Princeton University, Princeton, New Jersey 08544, USA}

%
%
\author{Arun Bansil}
\affiliation{Department of Physics, Northeastern University, Boston, Massachusetts 02115, USA}
\author{Guang Bian}\affiliation {Joseph Henry Laboratory, Department of Physics, Princeton University, Princeton, New Jersey 08544, USA}

\author{Hsin Lin}
\email{nilnish@gmail.com}
\affiliation{Centre for Advanced 2D Materials and Graphene Research Centre National University of Singapore, 6 Science Drive 2, Singapore 117546 and
Department of Physics, National University of Singapore, 2 Science Drive 3, Singapore 117542}
\author{M. Zahid Hasan \footnote{Corresponding authors (emails): mzhasan@princeton.edu and nilnish@gmail.com}}
\email{mzhasan@princeton.edu}
\affiliation {Joseph Henry Laboratory, Department of Physics, Princeton University, Princeton, New Jersey 08544, USA}

\pacs{}

\begin{abstract}
Relativistic fermions can be of three important varieties: Dirac, Majorana and Weyl. Recently, the Weyl semimetals, whose low-lying excitations are Weyl Fermions, have attracted worldwide attention due to their wide range of exotic electro-magnetic properties expected in theory. The experimental realization had remained elusive for a long time despite much efforts. Very recently, photoemission experiments (ARPES) have shown strong evidence identifying the first Weyl semimetal state in stoichiometric solid TaAs which marks the beginning of experimental research activity on this fascinating topic. In order to facilitate the transition of Weyl semimetals from a purely theoretical concept to the realm of experimental activities, it is of crucial importance to identify other material candidates. In this paper, we propose such a Weyl semimetal state in an inversion breaking, stoichiometric compound strontium silicide, SrSi$_2$, with many new and novel properties that are distinct from the TaAs family even in theoretical concepts. We theoretically show that SrSi$_2$ is a Weyl semimetal even without spin-orbit coupling and that, after the inclusion of spin-orbit coupling, two Weyl fermions stick together forming an exotic double Weyl fermion with quadratic dispersions and a higher chiral topological charge of 2. Moreover, we find that the Weyl nodes with opposite charges are located at different energies due to the absence of mirror symmetry in SrSi$_2$, leading to a unique topological quantum response that an external magnetic field can induce a dissipationless current. Our systematic results not only identify a much-needed robust Weyl semimetal candidate but also open the door to new topological Weyl physics that is not possible in the TaAs family of materials.
\end{abstract}
\date{\today}
\maketitle

Analogous to graphene and the three-dimensional topological insulator, Weyl semimetals are widely believed to open the next wave of experimental activities in topological condensed matter physics.\cite{Weyl, Nielsen1983, Balents_viewpoint, Wan2011, Murakami2007, TI_book_2014, Hasan2010, Haldane, FermiarcHasan, Singh2012, Ashvin_Review} A Weyl semimetal represents an elegant example of the correspondence between condensed matter and high energy physics, because its low energy excitations, the Weyl fermions, are massless particles that have played an important role in quantum field theory and the standard model but have not been observed as a fundamental particle in nature. A Weyl semimetal is also a topologically non-trivial metallic phase of matter extending the classification of topological phases beyond Kane-Mele topological insulators \cite{Wan2011, Murakami2007, TI_book_2014, Ashvin_Review}. The nontrivial topological nature guarantees the existence of exotic Fermi arc electron states on the surface of a Weyl semimetal. In contrast to a topological insulator where the bulk is gapped and only the Dirac cones on its surfaces are of interest, in a Weyl semimetal, both the Weyl fermions in the bulk and the Fermi arcs on the surface are fundamentally new and are expected to give rise to a wide range of exotic phenomena in bulk materials \cite{Nielsen1983, Zyuzin2012, SC, Ojanen, Carbotte2013, Potter2014, Parameswaran2014, Burkov2014}.

Very recently, photoemission experiments have reported strong evidence demonstrating the first realization of a Weyl semimetal state in the inversion symmetry breaking compound, TaAs \cite{Huang2015, Weng2015, TaAs Hasan, TaAs Ding}. Therefore, the field of Weyl semimetal is now at a stage that is similar to the first discovery of the topological insulator surface states in the BiSb semiconductors in 2007, meaning that it is crucially important to identify other robust Weyl semimetal candidates that possess properties that are distinct from the TaAs family of materials. In this paper, we propose a new type of Weyl semimetal in an inversion breaking, stoichiometric compound strontium silicide, SrSi$_2$. Our systematic first-principles band structure calculations show that SrSi$_2$ is a Weyl semimetal even in the absence of spin-orbit coupling. Upon including spin-orbit coupling, our study shows that the Weyl semimetal state remains intact, and, more interestingly, that two single Weyl fermions with the same chiral charge are bounded together forming a double Weyl fermion featuring a quadratic dispersion due to an additional $C_4$ symmetry. We find that such a  double Weyl fermion in SrSi$_2$ exhibits a high (larger than 1) chiral topological charge of 2. Our surface state calculations further show that the chiral charge 2 in the double Weyl fermions leads to an interesting phenomenon that two surface Fermi arcs thread through the same Weyl node. Furthermore, due to the simultaneous absence of mirror symmetry and inversion symmetry in SrSi$_2$, our calculations show that the Weyl nodes with opposite charges are located at different energies, leading to a new and unique topological transport phenomenon that an external magnetic field can induce an equilibrium dissipationless current \cite{Zyuzin2012}. Our prediction of the Weyl semimetal state in SrSi$_2$ serves as an important and timely contribution and pave the way for realizing many new phenomena such as quadratic Weyl fermions, higher chiral charges, magnetic field driven dissipationless currents, which are not possible in the known Weyl semimetal TaAs.

\begin{figure*}[tbp]
\begin{center}
\includegraphics[width=\textwidth] {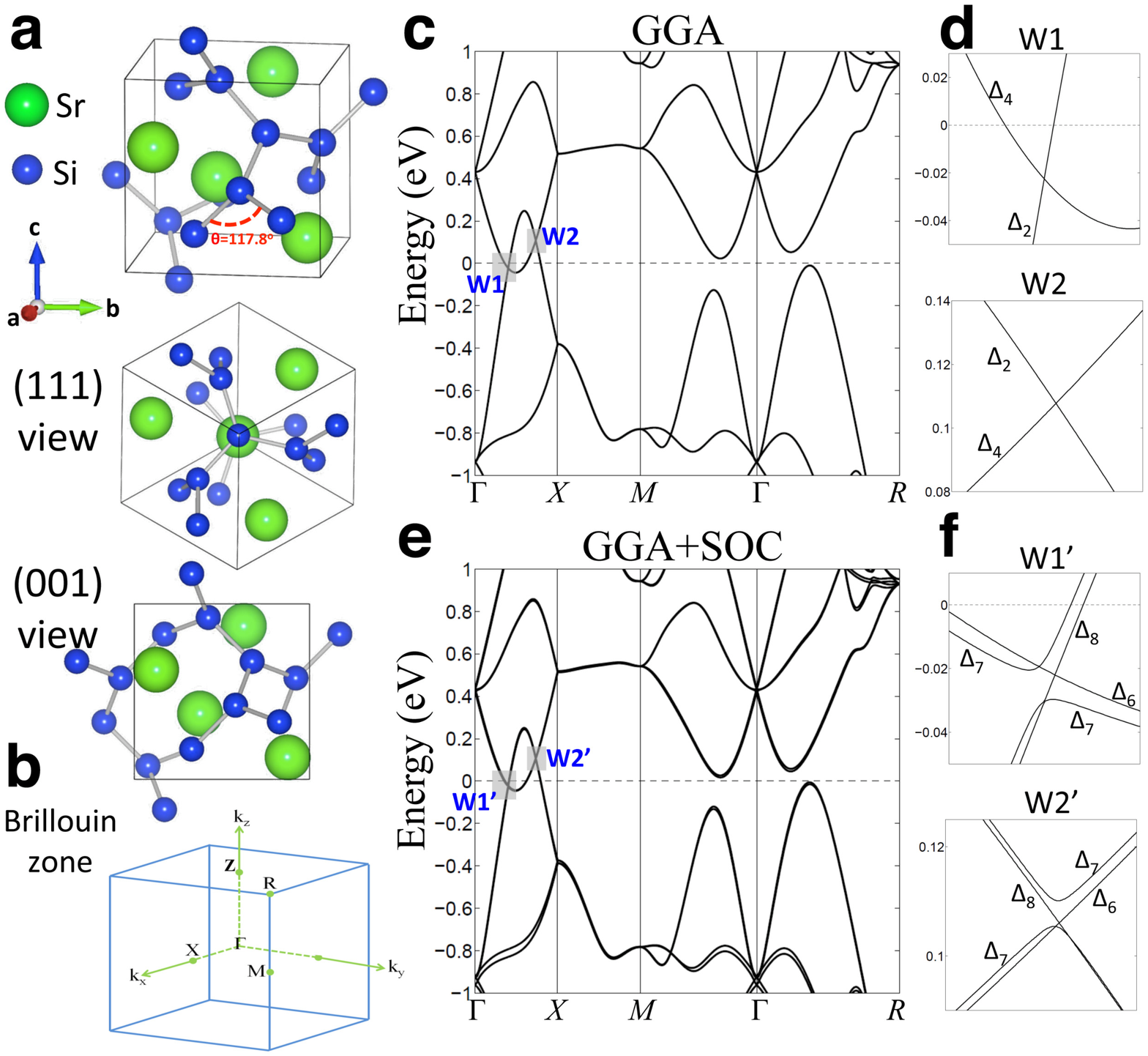}
\end{center}
\caption{\textbf{Crystal and electronic structure of SrSi}$_{2}$. \textbf{a,} Unit cell cubes from different views. A unit cell contains four Sr and eight Si atoms. \textbf{b,c,} Band structures along high-symmetry lines without and with SOC. In (b), the $\Delta _{2}$ and $\Delta _{4}$ bands will cross at W1 and W2 along $\Gamma $X, which are highlighted by gray bricks and brought into closeup in the right. The monopole charges at W1 and W2 are $C_{W}=+1$ and $-1$, respectively. In (c), SOC splits two bands: $\Delta_{2}\rightarrow \Delta _{7}+\Delta _{8}$ and $\Delta _{4}\rightarrow \Delta_{6}+\Delta _{7}$. Two $\Delta _{7}$ bands will repel by a gap, while $\Delta _{6}$ and $\Delta _{8}$ cross and create $C_{W}=\pm 2$ WPs, W1' and W2' respectively. There are other crossings between $\Delta _{7}$ and $\Delta _{8}$ and between $\Delta _{6}$ and $\Delta _{7}$, which are also WPs but will not be discussed here. }
\label{Fig1}
\end{figure*}

SrSi$_{2}$ crystalizes in a simple cubic lattice system. The lattice constant is $a=6.563$ $\textrm{\AA}^{-1}$ and the space group is $P4_{3}32$ (\#212). As seen in Fig.\ref{Fig1}\textbf{a}, the crystal lacks inversion and mirror symmetries. Since SrSi$_{2}$ is a nonmagnetic system that respects time-reversal symmetry, the absence of inversion symmetry is fundamental for realizing a Weyl semimetal phase. The lack of mirror symmetry also has important consequences to the energy positions of the Weyl nodes, which will be discussed in the following paragraphs. The bulk and (001) surface high symmetry points are noted in Fig.\ref{Fig1}\textbf{b}, where the centers of the square faces are the $X$ points, the centers of the edges are the $M$ points, and the corners of the cube are the $R$ points.

We try to understand the electronic properties of SrSi$_{2}$ at a qualitative level based on the ionic model. The electronic configuration of Sr is $4s^2$ whereas the electronic configuration of Si is $3s^23p^2$. Each Sr atom has a strong tendency to give out two electrons to achieve a full shell configuration in an ionic compound, leading to an ionic state of Sr$^{+2}$. This means that Si has an ionic state of Sr$^{-1}$ in SrSi$_{2}$, which is different from the most common ionic state of Si, Si$^{+4}$, as in SiO$_2$. This situation resembles another well-known semimetal Na$_3$Bi. In both compounds, an element that usually forms a positive ionic state (such as Si$^{+4}$ or Bi$^{+3}$) in an ionic compound is forced to form a negative ionic state (such as Si$^{-1}$ or Bi$^{-3}$). However, we emphasize that a key difference between SrSi$_{2}$ and Na$_3$Bi is that the SrSi$_{2}$ crystal breaks space inversion symmetry. Based on the above picture, we expect that the valence electronic states mainly arise from the $3p$ orbitals of Si. Indeed, this is confirmed by our first-principles calculation results. Fig.\ref{Fig1}\textbf{c} shows the calculated bulk band structure along high-symmetry directions in the absence of spin-orbit coupling. We observe a clear crossing between the bulk conduction and valence bands along the $\Gamma-X$ direction, which agrees with our expectation that SrSi$_{2}$ is likely to be a semimetal based on the ionic model picture. Interestingly, we note that the band crossing does not enclose any high symmetry or time-reversal invariant Kramers' points. In the vicinity of the crossings, the bands are found to disperse linearly along the $\Gamma-X$ direction as shown in Fig.\ref{Fig1}\textbf{d}. The two crossings along $\Gamma-X$ without spin-orbit coupling are noted as W$1$ and W$2$. At this point, one cannot conclude whether the conduction and valence bands between $\Gamma-X$ only touch at the two discrete points, W$1$ and W$2$ or if they dip into each other to form a 1D line node crossing. Since SrSi$_{2}$ lacks inversion symmetry, the spin degeneracy of the bulk bands is lifted except at the Kramers' points when spin-orbit coupling is turned on. However, remarkably, we find that the touchings at W$1$ and W$2$ remain intact (Fig.\ref{Fig1}\textbf{e,f}), where the band with one type of spin is gapped but the other spin remains gapless.

\begin{figure*}[tbp]
\begin{center}
\includegraphics[width=\textwidth] {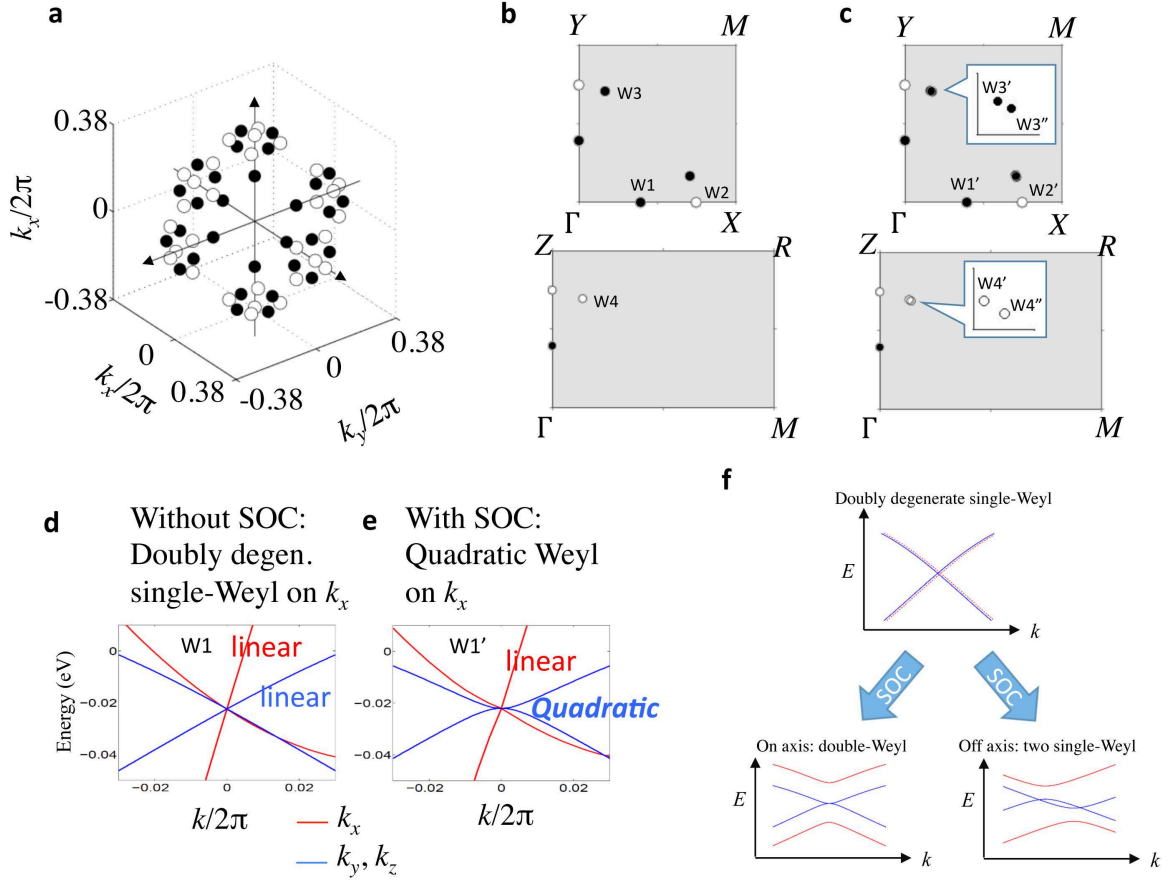}
\end{center}
\caption{\textbf{Distribution of Weyl points and their dispersions.} Without SOC 60 WPs in total, while with SOC 108 in total. \textbf{a,} Positions of WPs without SOC. Colors of circles stand for the signs of the monopole charge, black for positive and white for negative. \textbf{b,c,} Positions of WPs without (\textbf{b}) and with (\textbf{c}) SOC, respectively. Upper panels: the first quadrant of the $k_{z}=0$ plane; lower panels: the first quadrant of the $k_{x}-k_{y}=0$ plane. Distinct WPs not related by symmetry are labelled (W1$\sim$W4). When SOC is on, WPs will be tagged with prime or double prime superscripts. \textbf{d},\textbf{e}, Energy dispersions of WPs with and without SOC. \textbf{f}, The schematic of effects of SOC on the WPs. After hybridization of two degenerate single-WPs, those on the axes will double their charge, while those not on
the axes, instead, will split into two from each point.}
\label{Fig2}
\end{figure*}

We systematically study the nature of the band crossings in Fig.\ref{Fig2}. Let us first consider the band structure in the vicinity of the crossings W$1$ and W$2$ without spin-orbit coupling. Interestingly, we find that the bands disperse linearly along all three directions in vicinity of W$1$ (Fig.\ref{Fig2}\textbf{d}). This fact suggests that SrSi$_{2}$ is either a Dirac or a Weyl semimetal without spin-orbit coupling. We note that a Dirac semimetal is only possible in the presence of spin-orbit coupling \cite{Nagaosa}, and therefore the band crossings W$1$ and W$2$ are likely to be Weyl nodes. In order to rigorously show that this is indeed the case, we have checked the chiral charges of W$1$ and W$2$ by calculating the Berry flux through a closed surface that encloses W$1$ and W$2$. Our calculation shows that W$1$ and W$2$ carry nonzero a chiral charge, which proves that they are Weyl nodes. Therefore, SrSi$_{2}$ is already a Weyl semimetal even without spin-orbit coupling.

Now we study the band structure with spin-orbit coupling. This means that each Weyl node without spin-orbit coupling should be considered as two degenerate Weyl cones with the same chiral charge but the opposite physical spins. In general, spin-orbit coupling is expected to lift the spin degeneracy due to the lack of inversion symmetry in SrSi$_{2}$. For W$3$ and W$4$, as shown in Fig.\ref{Fig2}\textbf{c} the degeneracy of the physical spin of each W$3$ or W$4$ Weyl node is broken, and each W$3$ (W$4$) split into two Weyl nodes, W$3$' and W$3$'' (W$4$' and W$4$'') that are separated in momentum space. However, for W$1$ and W$2$, we observe a very interesting phenomenon: For example, if we consider a W$1$ that is located along the $k_x$ axis, then W$1$ does not split into two separated Weyl nodes as spin-orbit coupling is included. Instead, the dispersions of the Weyl bands along $k_y$ and $k_z$ directions become quadratic whereas the dispersion along $k_x$ remains linear, realizing a novel quadratic Weyl cone, which we note as W$1$' in Fig.\ref{Fig2}\textbf{c}. We have calculated the chiral charge associated with the quadratic Weyl node and our results reveal a chiral charge of 2. Usually, the Weyl cones in a Weyl semimetal disperse linearly and have chiral charges of $\pm1$. Therefore, a quadratic Weyl cone with a higher (larger than 1) chiral charge is already interesting and novel by itself. Furthermore, recent theories have suggested that a quadratic band touching in 3D may exhibit interesting non-Fermi liquid interaction effects because the long-range tail of the Coulomb repulsion is not screened \cite{QBT1, QBT2}. This may potentially lead to new correlated topological states. Another interesting property of SrSi$_2$ is that the Weyl nodes with opposite charges are located at different energies. This is due to the lack of mirror symmetry in the crystal, because a mirror symmetry operation would reflect a Weyl node on one side of the mirror plane to a Weyl node with the opposite chiral charge at the same energy. Such a property is interesting because It has been proposed that having Weyl nodes with the opposite chiralities at different energies can lead to a new and unique topological transport phenomenon that an external magnetic field can induce an equilibrium dissipationless current \cite{Zyuzin2012}.

\begin{figure*}[tbp]
\begin{center}
\includegraphics[width=\textwidth] {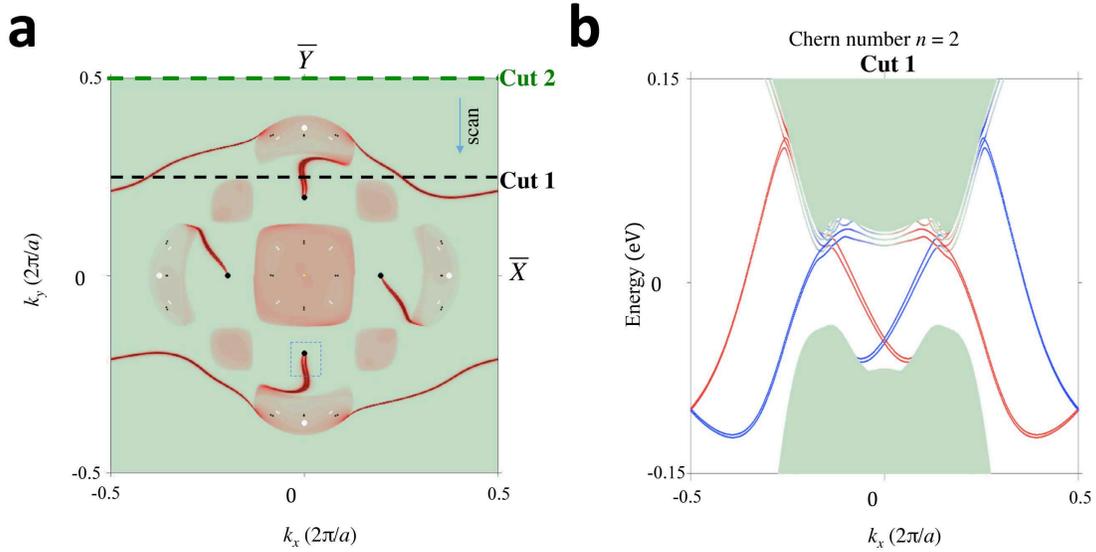}
\end{center}
\caption{\textbf{(001)-projected surface states on SrSi}$_{\mathbf{2}}$\textbf{.} \textbf{a,} Fermi arcs for top surface states at energy of W1' (double-WP). If zoom in, there are two lines with little separation owing to weak SOC. The Fermi arcs from the projection of a WP will merge into the bulk bands when the opposite charged WP is at different energy. \textbf{b,} $E$-$k$ dispersion at $k_{y}=\protect\pi /2$ (the horizontal dashed line in (\textbf{a})). Red lines for top surface states and blue ones for bottom surface states. For this $k_{y}=\protect\pi /2$ plane, it is a two-dimensional quantum Hall system with Chern number $-2$, such that there are two chiral edge states (for every surface) connecting the conduction and valence bands.}
\label{Fig3}
\end{figure*}

A key signature of a Weyl semimetal is the presence of Fermi arc surface states, which connect the Weyl points in a surface BZ. We present calculations of the (001) surface states in Fig.~\ref{Fig3}. The projected Weyl nodes are noted as black and white circles in Fig.~\ref{Fig3}\textbf{a}. The bigger circles correspond to the quadratic Weyl nodes with chiral charges of $\pm2$ whereas the smaller ones correspond to the linear Weyl nodes with chiral charges of $\pm1$. At first glance, it can be seen that the distribution of the bulk Weyl nodes is $C_4$ symmetric. By contrast, the Fermi surface of the surface states clearly violate the $C_4$ symmetry. This is consistent with the fact that the bulk crystal has $C_4$ symmetry but the (001) surface in fact violates such a symmetry. More importantly, Fig.~\ref{Fig3}\textbf{a} shows that the surface states are Fermi arcs because they connect the bulk Weyl nodes, as seen in the area that is highlighted by the blue dotted box. If one looks closely at the highlighted area, there are in fact two Fermi arcs that are very close to each other in momentum space. This is consistent with the chiral charge of 2 of the quadratic Weyl node that the Fermi arcs are terminated into. On the other hand, we note that because the quadratic Weyl nodes W$1$' and W$2$' are not located at the same energy, it is not possible to find an energy where only the Weyl nodes cross the Fermi level. The shaded areas in Fig.~\ref{Fig3}\textbf{a} represent the projections of the bulk bands that cross the Fermi level. Due to the existence of bulk Fermi surfaces, we find that the Fermi arcs may connect a Weyl node at one end but merge into the bulk continuum. We now study the energy dispersion of the surface states and the bulk band continuum. The calculated dispersion along Cut 1 (the black dotted line) is shown in Fig.~\ref{Fig3}\textbf{b}. The bulk band has a full gap along Cut 1. On the other hand, we find surface states that connect acres the bulk band gap. The red and blue lines show the surface states from the top and the bottom surface states. If one only counts the surface states from the top surface, then it can be seen that there are four right-going surface states and only two left-going ones. Therefore, the Chern number of this 2D $k$-slice, Cut 1, is 2. The nonzero Chern number is another key feature of a Weyl semimetal because in a topological insulator the Chern number of any 2D $k$-slice of the bulk BZ has to remain zero. We note that the Chern number of 2 is obtained from the electronic structure of the surface states (Fig.~\ref{Fig3}\textbf{b}) through the surface-bulk correspondence principle. On the other hand, we can also reach the same conclusion consistently by studying the bulk. If one studies the 2D $k$-slice Cut 2 (the green dotted line) that goes across the $\bar{Y}$ $(0, \pi)$ point (Fig.~\ref{Fig3}\textbf{a}), the Chern number of Cut 2 must be 0 because it goes through the Kramers points $\bar{Y}$ $(0, \pi)$ and $\bar{M}$ $(\pi, \pi)$. Then one can scan the 2D $k$-slice continuously from Cut 2 to Cut 1 along the direction indicated by the blue arrow in Fig.~\ref{Fig3}\textbf{a}, count the number of Weyl nodes and their chiral charges that have been crossed (Figs.~\ref{Fig2}\textbf{a-c}), and one will find that there are eight single Weyl nodes with the chiral charge of +1, eight single Weyl nodes with the chiral charge of -1 and one quadratic Weyl node with chiral charge of +2 that are scanned through during this process, which is entirely consistent with the Chern number of 2 of Cut 1.

\begin{figure*}[tbp]
\begin{center}
\includegraphics[width=\textwidth] {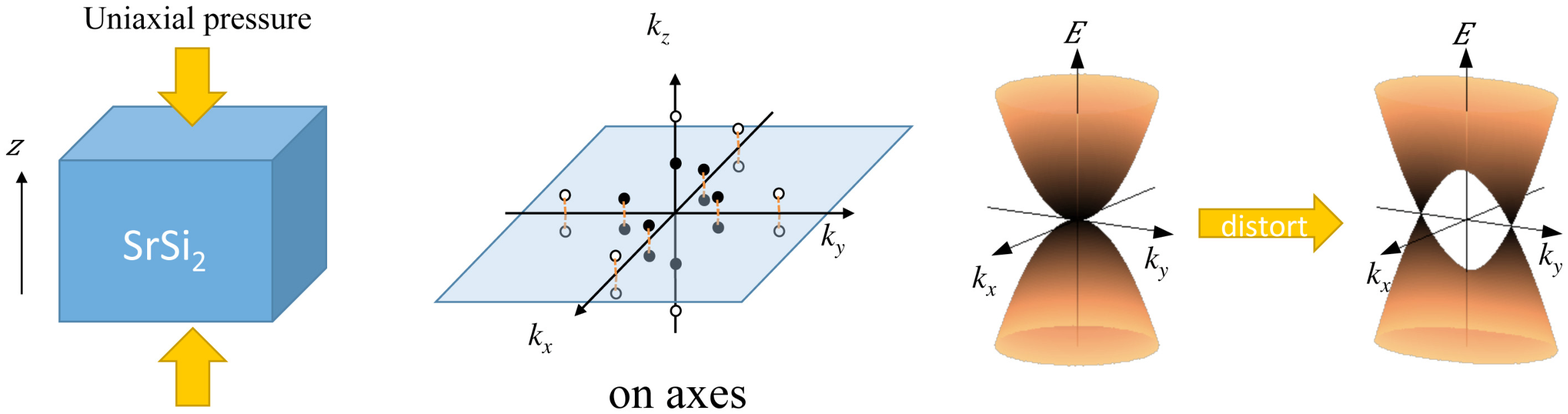}
\end{center}
\caption{\textbf{Quadratic Weyl points.} Under uniaxial pressure along the \textit{z} direction, $C_{4}$ rotation axes, $k_{x}$ and $k_{y}$, are lost, making double-WPs on these two axes split into two single-WPs away from the axes. The transformation of the energy dispersion is illustrated by cartoon on the most right.}
\label{Fig4}
\end{figure*}

Finally, we test the symmetry origin of the quadratic Weyl fermions in SrSi$_2$. We note that all the quadric Weyl nodes are located on the $k_x$, $k_y$ and $k_z$ axes which are the $C_4$ rotational axes of the crystal. This suggests that the quadratic Weyl fermions are likely protected by the $C_4$ symmetry. In order to test this speculation, we apply a uniaxial pressure along the $\hat{z}$ direction that compresses the lattice as shown in Fig.~\ref{Fig4}\textbf{a}. The consequence is that it breaks the $C_4$ symmetries along the $k_x$ and the $k_y$ directions but preserves the $C_4$ symmetry along the $k_z$ direction. We calculate the band structure under such a pressure. Our result in Fig.~\ref{Fig4}\textbf{b} shows that each quadratic Weyl node located on the $k_x$ or the $k_y$ axis is deformed into two single (linear) Weyl nodes, whereas all quadratic Weyl nodes on$k_z$ remain intact. Therefore, the quadratic Weyl node is protected by the $C_4$ rotational symmetry of the axis where the node is located. We note that previously quadratic Weyl fermions were only predicted in the Weyl semimetal candidate HgCr$_2$Se$_4$ \cite{Fang2012, Xu2011}. However, the experimental realization of the Weyl semimetal state in HgCr$_2$Se$_4$ has been proven to be difficult, because there is no preferred magnetization axis in its cubic structure, which likely leads to the formation of many small ferromagnetic domains. Here, our propose provides the first opportunity to realize the exotic quadric Weyl fermions in an inversion symmetry breaking single crystalline compound that does not rely on magnetic ordering.

Finally, we highlight the experimental feasibility of SrSi$_2$. We note that, prior to our prediction of the Weyl semimetal state in TaAs \cite{Huang2015}, there have been a number of candidates for a Weyl semimetal. However, those proposals have proven difficult to carry out because they rely on magnetic ordering to break time-reversal symmetry or fine-tuning the chemical composition of an alloy \cite{Wan2011, Xu2011, Burkov2011, Singh2012, HgCdTe, Liu2014}. Our proposal of TaAs was the first Weyl semimetal candidate in a stoichiometric, inversion symmetry breaking crystal, which does not rely on magnetic ordering over sufficiently large domains or fine-tuning the chemical composition  \cite{Huang2015}. The fact that the Weyl semimetal state has been observed in TaAs \cite{TaAs Hasan, TaAs Ding} demonstrates that such conditions are essential for the experimental realization. We emphasize that SrSi$_2$ is also a stoichiometric, inversion symmetry breaking single crystalline compound. This fact highlights the experimental feasibility for realizing the predicted Weyl semimetal state in SrSi$_2$.

In summary, we have proposed a new type of Weyl semimetal in SrSi$_2$. We have shown that SrSi$_2$ features many unusual properties not present in the TaAs family. SrSi$_2$ is a Weyl semimetal even in the absence of spin-orbit coupling. After including spin-orbit coupling, two single Weyl fermions with the same chiral charge are bounded together forming a double Weyl fermion featuring a quadratic dispersion due to an additional $C_4$ symmetry. We have found that such a double Weyl fermion in SrSi$_2$ exhibits a high (larger than 1) chiral charge of 2. Furthermore, the Weyl nodes with opposite charges are located at different energies, leading to a unique topological transport phenomenon that an external magnetic field can induce an equilibrium dissipationless current \cite{Zyuzin2012}. The fact that our proposal does not rely on magnetic ordering over sufficiently large domains or fine-tuning the chemical composition demonstrates its feasibility in experiments. 



\newpage

\clearpage

\clearpage

\clearpage


\begin{thebibliography}{99}

\bibitem{Weyl} Weyl, H. Elektron und gravitation. I. \textit{Z. Phys.} $\mathbf{56}$, 330-352 (1929).
\bibitem{Nielsen1983} Nielsen, H. B. \& Ninomiya, M.  The Adler-Bell-Jackiw anomaly and Weyl fermions in a crystal. \textit{Phys. Lett. B} \textbf{130}, 389 (1983).
\bibitem{Balents_viewpoint} Balents, L. Weyl electrons kiss. \textit{Physics} \textbf{4}, 36 (2011).
\bibitem{Wan2011} Wan, X. G., Turner, A. M., Vishwanath, A. \& Savrasov, S. Y. Topological semimetal and fermi-arc surface states in the electronic structure of pyrochlore iridates. \textit{Phys. Rev. B} \textbf{83}, 205101
(2011).
\bibitem{Murakami2007} Murakami, S. Phase transition between the quantum spin Hall and insulator phases in 3D: emergence of a topological gapless phase. \textit{New. J. Phys.} \textbf{9}, 356 (2007).
\bibitem{TI_book_2014} Hasan, M. Z., Xu, S.-Y. \& Neupane, M. Topological Insulators, Topological Crystalline Insulators, Topological Kondo Insulators, and Topological Semimetals. Preprint at http://arxiv.org/abs/1406.1040 (2014).
\bibitem{Hasan2010} Hasan, M.Z. \& Kane, C. L. Topological Insulators. \textit{Rev. Mod. Phys.} \textbf{82}, 3045 (2010).
\bibitem{FermiarcHasan} Xu, S.-Y., Liu, C., Kushwaha, S. \textit{et al.} Observation of "Fermi arcs" surface states in a topological metal.
    \textit{Science} \textbf{347}, 294 (2015). 
\bibitem{Haldane} Haldane, F. D. M. Attachment of surface "Fermi arcs" to the bulk Fermi surface: "Fermi-level plumbing" in topological metals. http://arxiv.org/abs/1401.0529 (2014).
  
\bibitem{Singh2012} Singh, B., Sharma, A., Lin, H., Hasan, M. Z., Prasad, R. \& Bansil, A. Topological electronic structure and Weyl semimetal in the TlBiSe$_2$ class of semiconductors. \textit{Phys. Rev. B} \textbf{86}, 115208 (2012).
\bibitem{Ashvin_Review} For a review, see, Turner, A. M. \& Vishwanath, A. Beyond band insulators: topology of semi-metals and interacting phases. Preprint at http://arxiv.org/abs/1301.0330 (2013).


\bibitem{Zyuzin2012} Zyuzin, A. A., Wu, S. \& Burkov, A. A. Weyl semimetal with broken time reversal and inversion symmetries. \textit{Phys. Rev. B} \textbf{85}, 165110 (2012).
\bibitem{SC} Cho, G. Y. \textit{et al.} Superconductivity of doped Weyl semimetals: Finite momentum pairing and electronic analog of the $^3$He-A phase. \textit{Phys. Rev. B} $\mathbf{86}$, 214514 (2012).
\bibitem{Ojanen} Ojanen, T. Helical Fermi arcs and surface states in time-reversal invariant Weyl semimetals. \textit{Phys. Rev. B} $\mathbf{87}$, 245112 (2013).
\bibitem{Carbotte2013} Ashby, P. E. C., Carbotte,J. P. Magneto-optical conductivity of Weyl semimetals. \textit{Phys. Rev. B} \textbf{87}, 245131 (2013).
\bibitem{Potter2014} Potter, A. C., Kimchi, I. \& Vishwanath, A. Quantum Oscillations from Surface Fermi-Arcs in Weyl and Dirac Semi-Metals. \textit{Nature Commun.} \textbf{5}, 5161 (2014).
\bibitem{Parameswaran2014} Parameswaran, S. A., Grover, T., Abanin, D. A., Pesin, D. A. \& Vishwanath A. Probing the Chiral Anomaly with Nonlocal Transport in Three-Dimensional Topological Semimetals. \textit{Phys. Rev. X} \textbf{4}, 031035 (2014).
\bibitem{Burkov2014} Panfilov, I., Burkov,A. A., Pesin,D. A. Density response in Weyl metals. \textit{Phys. Rev. B} \textbf{89}, 245103 (2014).
\bibitem{Huang2015} Huang, S. M., Xu, S. Y., Belopolski, I., Lee, C. C., Chang, G., Wang, B., Alidoust, N., Bian, G., Neupane, M., Bansil, A., Lin, H. \& Hasan, M. Z. An inversion breaking Weyl semimetal state in the TaAs material class. Preprint at http://arxiv.org/abs/1501.00755 (2015).
\bibitem{Weng2015} Weng, H. \textit{et al.} Weyl semimetal phase in non-centrosymmetric transition metal monophosphides. Preprint at http://arxiv.org/abs/1501.00060 (2015).

\bibitem{TaAs Hasan} Xu, S.-Y. \textit{et al.} Experimental realization of a topological Weyl semimetal phase with Fermi arc surface states in TaAs. Preprint at http://arxiv.org/abs/1502.03807 (2015).
\bibitem{TaAs Ding} Lv, B. Q. \textit{et al.} Discovery of Weyl semimetal TaAs. Preprint at http://arxiv.org/abs/1502.04684 (2015).

\bibitem{SrSi2 Crystal1} Kripyakevich, P.I. \& Gladyshevskii, E.I. The crystal structure of strontium disilicide. \textit{Kristallografiya} $\mathbf{11}$, 818-821 (1966).
\bibitem{SrSi2 Crystal2} Pringle, G.E. \textit{Acta Crystallographica B} $\mathbf{28}$, 2326-2328 (1972).
\bibitem{SrSi2 Crystal3} Evers, J. Transformation of three-dimensional three-connected silicon nets in SrSi$_2$. \textit{Journal of Solid State Chemistry} $\mathbf{24}$, 199-207, (1978).

\bibitem{Nagaosa} Yang, B.-J. \& Nagaosa, N. Classification of stable three-dimensional Dirac semimetals with nontrivial topology. \textit{Nat. Commun.} $\mathbf{5}$, 4898 (2014).

\bibitem{QBT1} Moon, E.-G., Xu, C., Kim, Y. B. \& Balents, L. Non-Fermi-Liquid and Topological States with Strong Spin-Orbit Coupling. \textit{Phys. Rev. Lett.} $\mathbf{111}$, 206401 (2013).


\bibitem{QBT2} Herbut, I. F. \& Janssen L. Topological Mott insulator in three-dimensional systems with quadratic band touching.  	\textit{Phys. Rev. Lett.} $\mathbf{113}$, 106401 (2014).


\bibitem{Fang2012} Fang, C., Gilbert, M. J., Dai, X. \& Bernevig, B. A. Multi-Weyl Topological Semimetals Stabilized by Point Group Symmetry. \textit{Phys. Rev. Lett.} \textbf{108}, 266802 (2012).

\bibitem{Xu2011} Xu, G., Weng, H., Wang, Z., Dai, X. \& Fang, Z. Chern Semimetal and the Quantized Anomalous Hall Effect in HgCr$_{2}$Se$_{4}$. \textit{Phys. Rev. Lett.} \textbf{107}, 186806 (2011).

\bibitem{Burkov2011} Burkov, A. A. \& Balents, L. Weyl semimetal in a topological insulator multilayer. \textit{Phys. Rev. Lett.} \textbf{107},127205 (2011).
\bibitem{HgCdTe} Bulmash, D., Liu, C.-X. \& Qi, X.-L. Prediction of a Weyl semimetal in Hg$_{1-x-y}$Cd$_x$Mn$_{y}$Te.\textit{Phys. Rev. B} $\mathbf{89}$, 081106(R) (2014).

\bibitem{Liu2014} Liu, J. P. \& Vanderbilt, D. Weyl semimetals from noncentrosymmetric topological insulators. \textit{Phys. Rev. B} \textbf{90}, 155316 (2014).

\end{thebibliography}
\end{document}